\documentclass[a4paper,11pt]{article}
\usepackage{amsmath}
\usepackage{amssymb}
\usepackage{epsfig}
\usepackage{graphics}

\begin{document}




\thispagestyle{empty}
\begin{titlepage}

\vspace*{1.0cm}
\vspace*{-1cm}
\hfill 
\parbox{3.5cm}{BUTP-99/50 \\ 
BUHE-99/05}\\ 

\vspace*{1.0cm}

\begin{center}
  {\large {\bf \hspace*{-0.2cm}
Glueball production in hadron and nucleus collisions
  } \footnote{
Contribution  to the International Europhysics Conference
High Energy 
Physics '99, Tampere, Finland, 15-21 July 1999, to appear in
the proceedings.}
\footnote{Work
      supported in part by the Schweizerischer Nationalfonds.}  }
  \vspace*{3.0cm} \\

{\bf

	\underline{S. Kabana}} \\
	Laboratory for High Energy Physics\\

	and\\

{\bf    P. Minkowski} \\
     Institute for Theoretical Physics \\

     University of Bern \\
     Sidlerstrasse 5 \\
     CH - 3012 Bern , Switzerland
   \vspace*{0.8cm} \\  
E-mails: {\tt Sonja.Kabana@cern.ch, mink@itp.unibe.ch} 
   \vspace*{0.8cm} \\

13. September 1999

\vspace*{2.0cm}

\begin{abstract}
The present discussion focusses on the
dominant production of the $0^{++}$ glueball in central collisions
in the
suppressed yet observable channel $0^{++} \rightarrow \pi^{+} \ \pi^{-}
\ \ell^{\ +} \ \ell^{-}$.

\end{abstract}
\end{center}
\end{titlepage}

\section{Introduction}

The assignement of the leading three binary glueballs 
 0$^{++}$(m = 1000 MeV,$\Gamma$ = 1000 MeV) (in \cite{PDG} $f_0$(400-1200)
and $f_0$(1370)), 
 0$^{-+}$(m = 1440 MeV,$\Gamma$ = 50-80 MeV) (in \cite{PDG} 
$\eta$(1440)) and
 2$^{++}$(m = 1710 MeV,$\Gamma$ = 133 MeV) (in \cite{PDG} $f_J$(1710))
has been proposed in \cite{PMWO,Ochstampere}.
In the present note we elaborate on the hypothesis that in high energy hadron
hadron and nucleus nucleus collisions the lowest mass glueballs
are copiously produced from the gluon rich environment, 
especially at high energy density.

\section{Central versus quasi-elastic production of glueball 0$^{++}$}

We take as a starting point the central production of 
glueball
$0^{++}$ in (1) $p p \rightarrow p X p$, $X= \pi^+ \pi^-$ \cite{akesson,PMWO} 
and (2) in two to two pseudoscalar meson scattering in the I = 0 channel
\cite{PMWO}.
The  large width of the glueball $0^{++}$  brings about
much different line shapes in (1) and (2) production channels.
At higher energy and energy density (reaction (1))
 the peak of $0^{++}$
 is shifted to the low mass region and the width is narrowed
as compared to reaction (2).
The suppressed channel $0^{++} \rightarrow \gamma \gamma$
is enforced by unitarity correction in reaction (2) where
 a branching fraction of 4-6 10$^{-6}$ was  found 
\cite{penington}.
In channel (1) however because of the narrower width the branching ratio
is expected to be larger than channel (2) 
by about a factor of 2. Therefore $\gamma \gamma$
and related channels for glueball 0$^{++}$ production become observable, 
while a quantitative theoretical determination is lacking.
The same is true for the other glueballs and illustrated in
the reaction
$\gamma^{*} \gamma^{*} \rightarrow f_J(1710) \rightarrow 
K \overline{K}$ \cite{L3}. 
The analysing power of the suppressed channel manifests itself in
the spin determination, 
which reveals a $\geq$ 70$\%$ spin 2 component \cite{L3_discussion}.
The exact determination of the  $\gamma \gamma $  width of $f_J$(1710)
demands control of the interference between $f_J$(1710) and $f'_2$(1525).
Notwithstanding
 the observed branching fraction of $f_J$(1710) $\rightarrow \gamma
\gamma$ does not contradict its glueball nature.

\section{Glueballs in central nucleus nucleus collisions}

We present a calculation of the production of
glueball $0^{++}$ dominating nucleus nucleus interactions
at midrapidity.
As outlined in the last section the mass of glueball $0^{++}$ 
is shifted to approximately 450 MeV in the high energy and
energy density environment.
On the purely hadronic level we interpret the onset of rising e.g. 
proton+proton total cross sections as implying gluon dominance with increasing
energy density. 
This brings us to our main hypothesis of enhanced glueball production 
in high energy density environment.
We raise the question whether 
measurement of glueball production reflects the number density of gluons
of their source.
Then an enhancement of the gluon degrees of freedom
measured through glueball production, identifies the presence of
 a quark gluon plasma phase \cite{qgp}.

\begin{figure}[htb]
\begin{center}
\vskip 1mm
\leavevmode
\resizebox{!}{16.0cm}{%
\includegraphics{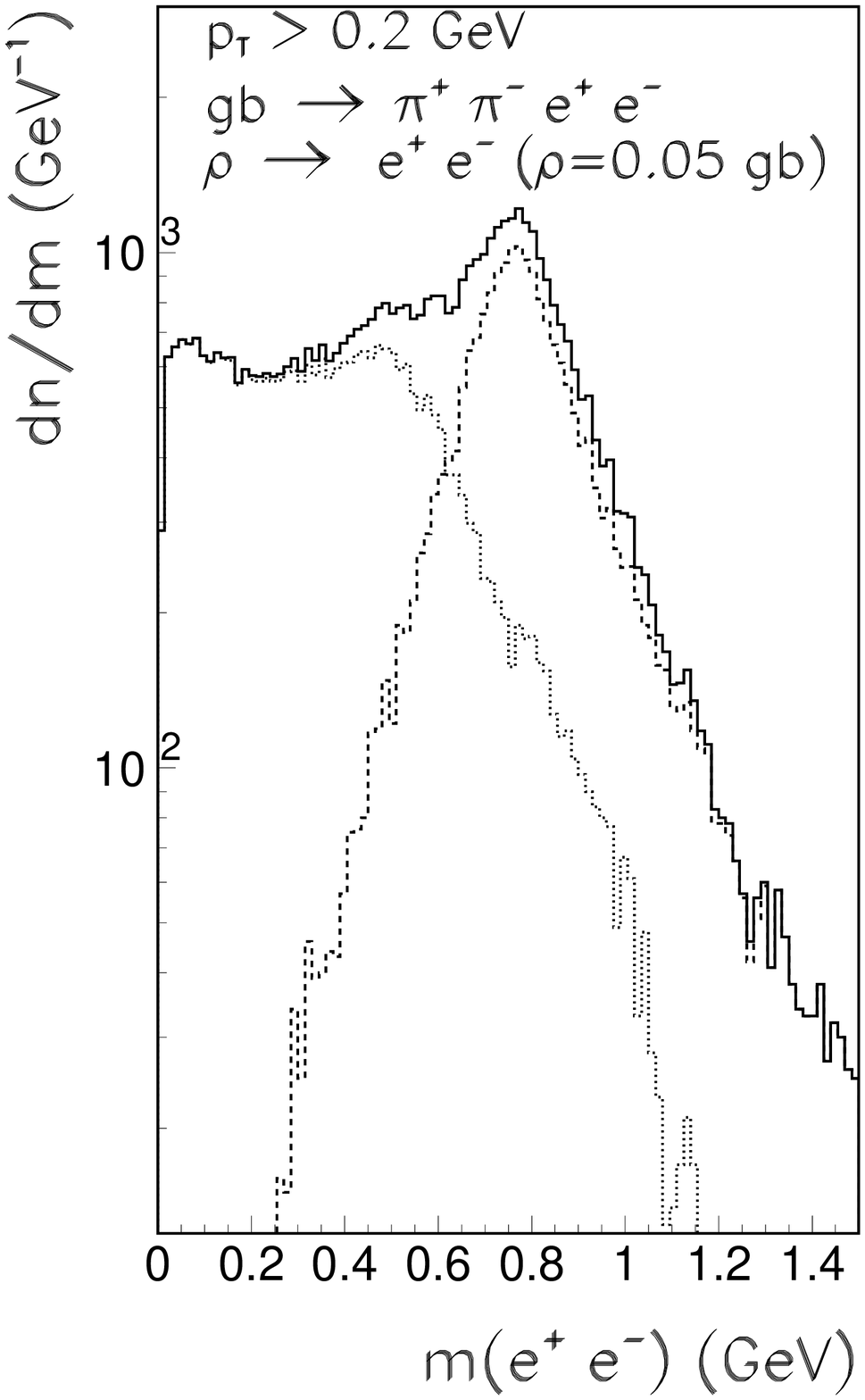}}
\vskip -2mm
\end{center}
{\center{
 \hspace*{1.0cm}
\begin{minipage}{13.cm}
\center{{\hspace*{-0.3cm} \mbox{ } 
}} 
\begin{flushleft}
\caption{
Invariant mass distribution of $e^+ e^-$ pairs
resulting from
 the assumed decay of the $0^{++}$ glueball
  state $0^{++} \rightarrow \pi^+ \pi^- e^+ e^- $
and from the decay $\rho \rightarrow e^+ e^- $.
}
 \label{gb_rho_nocuts}
\end{flushleft}
\end{minipage} }}
\end{figure}

We discuss the specific decay channel $0^{++} \rightarrow \pi^+ \pi^- 
e^+ e^-$ \cite{PMSK}.
The invariant mass distribution of $e^+ e^-$ pairs
produced in S+Au and Pb+Au reactions
at 200 and 158 GeV per nucleon respectively
shows a significant excess over expectations based on
sources of $e^+ e^-$ pair production reproducing proton+nucleus
collisions at the same energy \cite{na45}.
It is this excess we propose to explain:
\\
we assume that the production cross section times the branching
fraction of glueball $0^{++}$ in the above channel is in the ratio 20:1
relative to $\rho$ decaying into $e^+ e^-$.
We use the same mass distribution for $0^{++}$ as measured in
reference \cite{akesson}. The rapidity distribution is taken as a gaussian
with  standard deviation $\sigma$ of 0.6 units.
This assumption is supported by the measured y-distribution of
pion pairs resulting from the glueball $0^{++}$ decay shown in reference 
\cite{akesson}.

For the transverse mass distribution of the glueball we choose
 an exponential shape with inverse slope of 150 MeV.
The choice of this shape is justified by the measured shape
of the transverse mass distribution of most hadrons produced in
Pb Pb collisions at 158 A GeV with a slope seen to increase
with decreasing mass of the particles due to presence of a collective
transverse flow  \cite{antinori_qm99}.
If the glueball 0$^{++}$ were the dominant source of pions one
would expect it to have smaller inverse slope than its decay products.
Therefore the glueball is expected to lye outside the systematics
of inverse slopes as a function of mass of several measured
quark states produced in Pb Pb collisions \cite{antinori_qm99}.
Pions have an inverse slope between 150-210 MeV \cite{antinori_qm99}.
 Hence we take for the glueball 0$^{++}$ the smallest measured
 transverse mass slope of  pions: 150 MeV.
The $\rho$ meson was assumed to have the same rapidity distribution as the
glueball $0^{++}$ and exponential transverse mass distribution
with inverse slope  200 MeV.
 
\noindent
In addition, in order to simulate the experimental response,
we introduced an error of $\Delta p /p$ = 5$\%$  for the momenta
of the decay products $e^+ e^-$.
 
\noindent
Figure \ref{gb_rho_nocuts} shows
the invariant mass distribution of $e^+ e^-$ pairs
resulting from the assumed copious production and decay 
of the $0^{++}$ glueball state : 
$ \ \ 0^{++} \  \rightarrow \pi^+ \pi^- e^+ e^- $
and from the decay 
$ \ \rho \ \rightarrow e^+ e^- $.

\noindent
  A cut on the transverse momentum of the lepton pair
  of 0.2 GeV is imposed, as it was applied to the experimental
data  \cite{na45}.
The $e^+ e^-$ pairs from the decay of the $0^{++}$ glueball
state significantly populate the region of invariant mass where the excess is
seen by the NA45 experiment.

\section{Conclusions}

We have shown that dominant glueball production
can account for the observed $e^+ e^-$ invariant mass distribution
in central Pb+Au and S+Au collisions at 158 and 200 GeV
per nucleon respectively.
We expect that also glueballs $0^{-+}$ and $2^{++}$ are copiously produced
in  gluon rich environment.
The experimental verification of this hypothesis 
in channels like: $\pi \pi$, $K \overline{K}$, $K \overline{K} \gamma$
and $\gamma \gamma$ remains to be achieved.

\end{document}